\documentclass[journal]{IEEEtran}

\usepackage[colorlinks]{hyperref}

\usepackage{booktabs}
\usepackage{tabularx}

\usepackage{amssymb}
\usepackage{amsbsy}
\usepackage{amsmath}
\usepackage{bm}
\usepackage{verbatim}
\usepackage{mathrsfs}
\usepackage{amsfonts}
\usepackage{graphicx}
\usepackage[tight,footnotesize]{subfigure}
\usepackage[10pt]{moresize}
\usepackage{array}
\usepackage{color}
\usepackage{epsfig}
\usepackage{stfloats}
\usepackage{balance}

\usepackage{hyperref} 

\usepackage{amssymb}

\usepackage[noend]{algpseudocode}
\usepackage{algorithmicx,algorithm}

\usepackage{setspace}
\usepackage{cases}

\usepackage{graphicx}
\usepackage{epstopdf}
\usepackage{multirow}
\usepackage{extarrows}

\usepackage{pifont} 
\usepackage{enumerate}
\usepackage{enumitem}

\newcommand{\subparagraph}{}
\usepackage{titlesec}
\titlespacing{\section}{0pt}{2 ex plus .0ex minus .0ex}{1ex plus .0ex}
\titlespacing{\subsection}{0pt}{1.5 ex plus .0ex minus .0ex}{0.8 ex plus 0.0ex}
\titlespacing{\subsubsection}{0pt}{0.5ex plus .0ex minus .0ex}{0.0ex plus .0ex}

\usepackage{amsmath} 
\allowdisplaybreaks[4]

\ifCLASSINFOpdf

\else

\fi

\begin{document}
%
\title{Toward Practical Fluid Antenna Systems:\\Co-Optimizing Hardware and Software for\\ Port Selection and Beamforming}
\author{Sai Xu,~\IEEEmembership{Member,~IEEE}, 
            Kai-Kit Wong,~\IEEEmembership{Fellow,~IEEE}, 
            Yanan~Du,~\IEEEmembership{Member,~IEEE},
            Hanjiang Hong,~\IEEEmembership{Member,~IEEE}, 
            Chan-Byoung Chae, \emph{Fellow, IEEE}, 
            Baiyang Liu,~\IEEEmembership{Senior Member,~IEEE}, and 
            Kin-Fai Tong, \emph{Fellow, IEEE}
\vspace{-3mm}

\thanks{The work of S. Xu, K. K. Wong and H. Hong is supported by the Engineering and Physical Sciences Research Council (EPSRC) under Grant EP/W026813/1.}
\thanks{The work of H. Hong is supported by the Outstanding Doctoral Graduates Development Scholarship of Shanghai Jiao Tong University.}
\thanks{The work of C.-B. Chae was in part supported by the Institute for Information and Communication Technology Planning and Evaluation (IITP)/NRF grant funded by the Ministry of Science and ICT (MSIT), South Korea, under Grant RS-2024-00428780 and 2022R1A5A1027646.}
\thanks{The work of K. F. Tong and B. Liu was funded by the Hong Kong Metropolitan University, Staff Research Startup Fund: FRSF/2024/03.}

\thanks{S. Xu, K. Wong and H. Hong are with the Department of Electronic and Electrical Engineering, University College London, WC1E 7JE, London, UK. K. Wong is also with Yonsei Frontier Lab, Yonsei University, Seoul, South Korea (e-mail: $\rm\{sai.xu,kai\text{-}kit.wong,hanjiang.hong\}@ucl.ac.uk$).}
\thanks{Y. Du is with the Department of Electronic and Electrical Engineering, University of Sheffield, S1 4ET, UK (e-mail: $\rm yanan.du@sheffield.ac.uk$).}
\thanks{C.-B. Chae is with the School of Integrated Technology, Yonsei University, Seoul, 03722 South Korea (e-mail: $\rm cbchae@yonsei.ac.kr$).}
\thanks{B. Liu and K. F. Tong are with the School of Science and Technology, Hong Kong Metropolitan University, Hong Kong SAR, China (e-mail: $\rm \{byliu, ktong\}@hkmu.edu.hk$).}

\thanks{Corresponding author: Kai-Kit Wong.}
}

\maketitle
\begin{abstract}
This paper proposes a hardware-software co-design approach to efficiently optimize beamforming and port selection in fluid antenna systems (FASs). To begin with, a fluid-antenna (FA)-enabled downlink multi-cell multiple-input multiple-output (MIMO) network is modeled, and a weighted sum-rate (WSR) maximization problem is formulated. Second, a method that integrates graph neural networks (GNNs) with random port selection (RPS) is proposed to jointly optimize beamforming and port selection, while also assessing the benefits and limitations of random selection. Third, an instruction-driven deep learning accelerator based on a field-programmable gate array (FPGA) is developed to minimize inference latency. To further enhance efficiency, a scheduling algorithm is introduced to reduce redundant computations and minimize the idle time of computing cores. Simulation results demonstrate that the proposed GNN-RPS approach achieves competitive communication performance. Furthermore, experimental evaluations indicate that the FPGA-based accelerator maintains low latency while simultaneously executing beamforming inference for multiple port selections.
\end{abstract}
\IEEEpeerreviewmaketitle
\section{Introduction}
\IEEEPARstart{F}LUID antenna systems (FASs) have emerged as a novel concept for enhancing wireless communications~\cite{New2025Tutorial}. A fluid antenna (FA) is a fluidic, conductive, or dielectric structure that can dynamically adjust its shape and position, referred to as a \textit{port}, to modify key radio-frequency (RF) characteristics through software control \cite{Wong2020Fluid, Wong2022Bruce}. Unlike fixed-position antennas~\cite{Sanayei2004Antenna, MIMO2004Molisch}, the flexibility of FA allows for adjustment of dimensions and orientation of radiating elements, as well as directional beamforming without requiring analog or digital signal processing~\cite{New2025Tutorial}. Originally proposed by Wong {\em et al.}~in \cite{I22_wong2020perflim,I20_wong2021FAS}, FAS has since become a reality, with prototypes and experimental results reported in \cite{I24_shen2024design,I26_zhang2024pixel,Liu-2025arxiv}. Recently, Lu {\em et al.}~provided an electromagnetic perspective of FAS in \cite{Lu-2025}.

Despite its potential, a fundamental challenge arises. 

\vspace{2mm}
\begin{quote}
\begin{center}
{\em How can beamforming and port selection in FASs be efficiently optimized in practice?}
\end{center}
\end{quote}
\vspace{2mm}

To tackle this challenge, recent studies have explored mathematical optimization~\cite{Chen2025Joint, Zhou2024Fluid, Qin2024Antenna, Zou2024Shifting, Efrem2024Transmit} and learning-based approaches~\cite{ Wang2024Learning} for beamforming and port selection in FASs, as detailed in Section \ref{ssec:jopt}. Although these optimization algorithms achieve exceptional communication performance, their high computational complexity limits their applicability in real-time scenarios. Hence, it is necessary to seek a trade-off between communication performance and computational latency. Existing algorithmic studies often overlook hardware implementation, which further limits execution efficiency.

Motivated by this, this paper develops a hardware-software co-design framework for FAS. Specifically, an FA-enabled downlink multi-cell multiple-input multiple-output (MIMO) network is considered. Unlike existing methods, the proposed approach integrates graph neural networks (GNNs) with random port selection (RPS) to jointly optimize beamforming and port selection. Furthermore, an instruction-driven deep learning accelerator based on a field-programmable gate array (FPGA) is designed, featuring an efficient scheduling algorithm tailored to the proposed GNN-RPS approach. The main contributions of this work are summarized as follows:
\begin{itemize}
\item A GNN-RPS approach is developed to jointly optimize beamforming and port selection in FAS. The proposed neural network employs a multi-GNN architecture with centralized training and distributed deployment. Simulation results show that the GNN-based scheme outperforms traditional methods such as minimum mean square error (MMSE), zero-forcing (ZF), and maximum ratio transmission (MRT), while also providing performance insights into the use of RPS.

\item An instruction-driven FPGA-based accelerator is designed with a customized instruction set architecture (ISA) and a specialized overlay micro-architecture, both tailored to the proposed GNN-RPS optimization. The accelerator incorporates both software and hardware optimizations to enhance parallelism and reduce computational latency. Additionally, it exhibits scalability and flexibility, allowing future upgrades or reconfiguration as the GNN algorithm evolves.

\item An efficient scheduling algorithm is proposed to concurrently execute multiple beamforming inference tasks with different port selections on the FPGA. To address the memory-intensive nature of GNN inference and reduce idle time of computing cores, the algorithm leverages model parameter sharing, intermediate result reuse, and computation reordering to minimize memory access and eliminate redundant operations. Concurrent task execution maximizes core utilization and overall throughput.
\end{itemize}

The remainder of this paper is organized as follows. Section~\ref{sec:related} reviews related work relevant to this research. Section~\ref{sec:model} then presents the system model and formulates the optimization problem. Section~\ref{sec:beamport} introduces the GNN-RPS approach. Section \ref{sec:fpga} details the design of the instruction-driven FPGA-based accelerator. Section~\ref{sec:evaluate} presents both simulation and experimental results. Section~\ref{sec:conclude} concludes the paper.

\section{Related Work}\label{sec:related}
This research spans three major areas: (1) beamforming and port selection in FASs, (2) GNN-based beamforming optimization, and (3) FPGA-based deep learning acceleration. Related work in each of these areas is reviewed below.

\subsection{Beamforming and Port Selection in FASs}\label{ssec:jopt}
FAS-related optimization problems often take the form of mixed-integer programs, when the ports are modeled as discrete variables. To solve such problems, Chen \textit{et al.}~\cite{Chen2025Joint} proposed an alternating optimization (AO) approach to maximize energy efficiency in near-field communication by iteratively optimizing beamforming and FA positioning. Following similar principles, Zhou \textit{et al.}~\cite{Zhou2024Fluid} maximized downlink communication performance using AO while ensuring sensing and power constraints. Qin \textit{et al.}~\cite{Qin2024Antenna} applied AO with a penalty method and successive convex approximation to jointly optimize FA positions and beamforming in multiple-input single-output (MISO) downlink systems, aiming to minimize total transmit power. In addition to AO algorithms, Zou \textit{et al.}~\cite{Zou2024Shifting} developed an iterative algorithm integrating sparse optimization, convex approximation, and penalty methods to minimize transmit power while satisfying sensing and communication requirements. Efrem \textit{et al.}~\cite{Efrem2024Transmit} introduced joint convex relaxation with reduced exhaustive search to enhance fluid-MIMO channel capacity by optimizing antenna port positions. Wang \textit{et al.}~\cite{Wang2024Learning} introduced a deep reinforcement learning approach to maximize sum-rate while meeting sensing constraints. Although existing beamforming and port selection algorithms effectively address the optimization problems within their respective scenarios, they often suffer from high computational complexity and overlook hardware execution efficiency.

\subsection{GNN-based Beamforming Optimization}\label{ssec:gnn}
A GNN is a well-known neural network architecture specifically designed for graph-structured data, which learns the relationships between graph elements by propagating and aggregating information through the nodes and edges. Recently, GNN-based optimization has gained attention for beamforming design in wireless communication systems. Jiang \textit{et al.}~\cite{Jiang2021Learning} employed GNN to predict beamforming vectors from received pilots and user locations. Chen \textit{et al.}~\cite{Chen2024Distributed} proposed a multi-GNN architecture to maximize the weighted sum-rate (WSR) for reconfigurable intelligent surface (RIS) assisted cell-free MIMO networks. Xu \textit{et al.}~\cite{Xu2025TVT} investigated the GNN-based hybrid beamforming for satellite-terrestrial communications. Moreover, Li \textit{et al.}~\cite{Li2024GNN} utilized GNNs for sum-rate maximization in multiuser MISO networks, optimizing beamforming vectors under the data rate requirements of users and the power constraints of the base station (BS). After that, Mishra \textit{et al.}~\cite{Mishra2024Graph} proposed a GNN-based power allocation strategy, which leveraged the geometric structure of partially connected cell-free massive MIMO systems to maximize the minimum signal-to-interference-plus-noise ratio (SINR). Li \textit{et al.}~\cite{Li2024MaxMin} explored complex edge graph attention networks to optimize max-min fairness and handle power budget constraints. Xu \textit{et al.}~\cite{Xu2023Distributed} also devised a distributed auto-learning GNN, which enabled efficient interference mitigation and beamforming in a multi-cell cluster-free non-orthogonal multiple access system, and reduced overhead while improving sum rate performance. Despite its promise, the application of GNN to beamforming and port selection optimization remains unexplored.

\subsection{FPGA-based Deep Learning Acceleration}\label{ssec:fpga}
FPGA is commonly used to accelerate deep neural network (DNN) inference, thanks to its reconfigurability, energy efficiency, and fast development cycles, making it particularly well-suited for real-time and power-efficient applications~\cite{Xu2015GSM}. Typically, two architectural paradigms are considered: a fully pipelined architecture~\cite{Wei2018TGPA, Zhang2020DNNExplorer} and a non-pipelined one~\cite{Genc2021Gemmini, Yu2019OPU, Yu2020Light-OPU, Tang2024Grapht-OPU}. The former handles layers in sequence to enhance resource utilization, whereas the latter approach utilizes a universal compute unit for all the DNN layers. In terms of the fully pipelined architecture, Wei \textit{et al.}~\cite{Wei2018TGPA} proposed a tile-grained pipeline architecture that can reduce latency and improve performance for DNN inference by enabling pipelined execution of multiple tiles within a single input image across heterogeneous accelerators. Then Zhang \textit{et al.}~\cite{Zhang2020DNNExplorer} introduced a novel FPGA-based DNN accelerator design paradigm and automation tool, enabling fast exploration and optimization of accelerator architectures for various DNN networks. 

In terms of the non-pipelined architecture, Genc \textit{et al.}~\cite{Genc2021Gemmini} introduced an open-source DNN accelerator generator that designed efficient application-specific integrated circuit accelerators while accounting for system-level effects like resource contention and operating system overheads. Yu \textit{et al.}~\cite{Yu2019OPU, Yu2020Light-OPU} proposed a domain-specific FPGA overlay processor to accelerate DNN networks by providing software-like programmability. Based on~\cite{Yu2019OPU, Yu2020Light-OPU}, Tang \textit{et al.}~\cite{Tang2024Grapht-OPU} applied the same methodology to the acceleration of GNN. While existing FPGA-based deep learning accelerators can execute the GNN-RPS approach for beamforming and port optimization, they did not fully take into account the approach's unique characteristics, thereby limiting performance potential.

\section{System Model and Problem Formulation}\label{sec:model}
\subsection{System Model}\label{ssec:model}
\begin{figure}
\centering
\includegraphics[width=\linewidth]{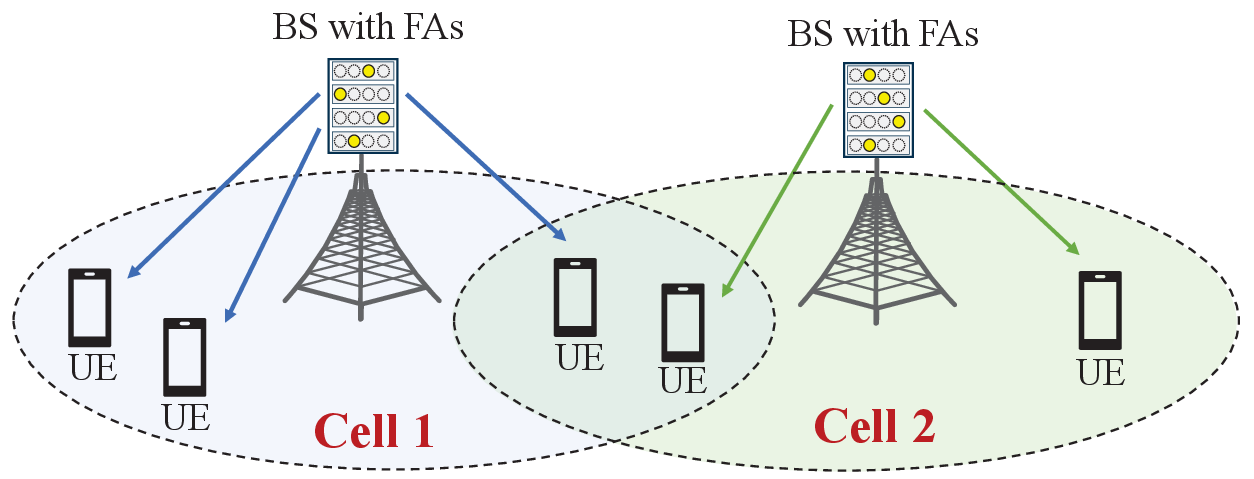}
\caption{An FA-enhanced downlink multi-cell MIMO network, where each BS with FAs simultaneously transmits multiple data streams to its associated users each equipped with a fixed-position antenna in a cell.}\label{Fig1}
\end{figure}

Fig.~\ref{Fig1} depicts an FA-enhanced downlink multi-cell MIMO network where each BS with FAs simultaneously transmits multiple data streams to its associated user equipments (UEs) each with a fixed-position antenna in a cell. Specifically, the network is composed of $I$ cells, in which $K_i$ UEs are served in the $i$-th cell for $i \in \mathcal{I} = \{1, 2, \dots, I\}$. In terms of hardware configuration, each BS is equipped with $N$ FAs, each of which has an RF chain and $L$ ports. As the RF chain is switchable among different ports, the spatial degrees of freedom (DoF) between BSs and UEs are higher than those with traditional fixed-position antennas, allowing for flexible configuration of channel conditions.
 Additionally, the FAs are spaced far enough apart (typically $\geq \lambda/2$) to ensure negligible spatial correlation and mutual coupling, where $\lambda$ denotes the wavelength. For each FA, the ports are uniformly distributed along a linear dimension of length $W \lambda$.\par

In contrast to the negligible correlation between FAs, there can be considerable spatial correlation among the ports within a given FA. Assuming rich scattering, the spatial correlation can be characterized by Jake's model~\cite{New2024information}, expressed as
\begin{align}
J_{(l, l')}  = \delta^2 J_0 \left( \frac{2 \pi W |l - l'|}{L-1} \right),
\end{align}
where $J_0 (\cdot)$ is the zero-order Bessel function of the first kind, $\delta^2$ denotes the path loss of the channel. Through eigenvalue decomposition, the spatial correlation matrix $\textbf{J}$, whose $(l, l')$-th element is denoted as $J_{(l, l')}$, can be represented by 
\begin{align}
\textbf{J}  = \textbf{Q} \boldsymbol{\Lambda} \textbf{Q}^H,
\end{align}
where $\textbf{Q} \in {\mathbb{C}^{L \times L}} $ contains the eigenvectors of $\textbf{J}  \in {\mathbb{C}^{L \times L}} $, and $\boldsymbol{\Lambda}  \in {\mathbb{C}^{L \times L}} $ is a diagonal matrix of the corresponding eigenvalues. Based on this, the channel vector from all the ports of an FA to a UE can be written as
\begin{align}
\bar{\textbf{h}}  = \textbf{Q} \boldsymbol{\Lambda}^{\frac{1}{2}} \textbf{z}, \label{E1} 
\end{align}
where $\textbf{z}$ is an independent and identically distributed (i.i.d.) complex Gaussian random vector, i.e., $\textbf{z} \sim \mathcal{CN}(\textbf{0}, \textbf{I}_L)$.

\subsection{Problem Formulation}\label{ssec:problem}
Let $i \in \mathcal{I} = \{1, 2, \dots, I\}$, $\mathcal{K}_i = \{1, 2, \dots, K_i\} $, $\mathcal{N} = \{1, 2, \dots, N\}$, and $\mathcal{L} = \{1, 2, \dots, L\} $ denote the sets of cells, UEs in the  $i$-th cell, FAs at each BS, and ports of each FA, respectively. Based on the spatial correlation model in \eqref{E1}, the channel vector from the $j$-th BS to the $k$-th UE in the $i$-th cell, denoted by $\textbf{h}_{ik,j} \in {\mathbb{C}^{N \times 1}}$, comprises $N$ independent entries, each having $L$ correlated options. Thus, $\textbf{h}_{ik,j}$ has a vast number of possible combinations, with different combinations corresponding to varying channel condition configurations. We assume that all the involved channels experience flat fading with fully available channel state information (CSI).

Taking into account the narrowband transmission during a given time slot, the received signal at the $k$-th UE in the $i$-th cell is given by
\begin{multline}
y_{ik}=\textbf{h}_{ik,i} \textbf{w}_{ik} x_{ik}  +
\underbrace{ \sum_{r \in \mathcal{K}_i\atop r \neq k} \textbf{h}_{ik,i} \textbf{w}_{ir} x_{ir}}_\text{Intra-cell interference} +\\
 \underbrace{\sum_{j \in \mathcal{I}, r \in \mathcal{K}_j\atop j \neq i} \textbf{h}_{ik,j}  \textbf{w}_{jr} x_{jr}}_\text{Inter-cell interference} +  n_{ik},
\end{multline}
where $\textbf{w}_{jr} \in {\mathbb{C}^{N \times 1}}$ denotes beamforming vector for the $r$-th data stream at the $j$-th BS. Consequently, the achievable communication rate at the $k$-th UE in the $i$-th cell is given by
\begin{align}
R_{ik} &=  \log_2\left( 1 + \frac{ | \textbf{h}_{ik,i}^H  \textbf{w}_{ik} |^2}{ \sum_{(j, r) \neq  (i, k)}  | \textbf{h}_{ik,j}  \textbf{w}_{jr} |^2 +\sigma^2} \right).
\end{align}
The objective of this paper is to maximize the WSR subject to the transmit power constraint in the considered FA-enhanced downlink multi-cell MIMO network. The optimization problem is formulated as
\begin{align*}
(\text{P1}) \quad \underset{\textbf{w}_{ik} \in \mathcal{W}(\textbf{s})} \max \quad &  \sum_{(i, k)} \omega_{ik} R_{ik}, \\
\text{s.t.} \quad
 & \text{C1}: \text{Tr} \left( \sum_{k \in \mathcal{K}_i }  \textbf{w}_{ik} \textbf{w}_{ik}^H \right) \leq P, ~i \in \mathcal{I}, \\  
 & \text{C2}: \textbf{s} \in \mathcal{S},
\end{align*}
where $\mathcal{W}(\textbf{s})$ and $\mathcal{S}$ denote the collections of $\textbf{w}_{ik}$ and all the combinations for  port selection, respectively, $\textbf{s}$ is a possible port selection and $P$ is the power budget at each BS.

\section{Beamforming and Port Selection}\label{sec:beamport}
The optimization problem (P1) requires joint design of beamforming and port selection, resulting in a mixed-integer non-convex formulation that is computationally prohibitive. To overcome this challenge, in this section, we present a GNN-RPS approach to optimize beamforming and port selection for the considered system, including GNN-based beamforming optimization and port selection schemes.

\subsection{GNN-Based Beamforming Optimization}\label{ssec:gnn-opt}
\begin{figure}
\centering
\includegraphics[width=3.2in]{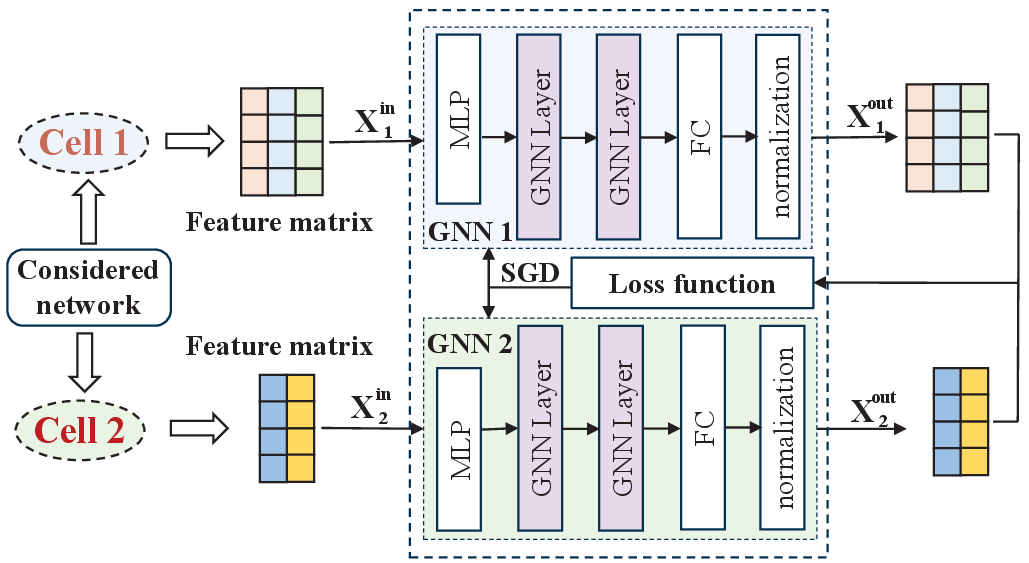}
\caption{An illustration of graph representation and neural network architecture.}\label{Fig2}
\end{figure}

For a given port selection, the non-convex problem (P1) can be simplified to
\begin{align*}
(\text{P2}) \quad \underset{\textbf{w}_{ik} \in \mathcal{W}} \max \quad &  \sum_{(i, k)} \omega_{ik} R_{ik},  \\
\text{s.t.} \quad
 & \text{C1}: \text{Tr} \left( \sum_{k \in \mathcal{K}_i }  \textbf{w}_{ik} \textbf{w}_{ik}^H \right) \leq P, ~i \in \mathcal{I}, 
\end{align*}
where $\mathcal{W}$ is the collection of $\textbf{w}_{ik}$ for a given port selection. Nevertheless, the optimization problem (P2) still remains challenging to solve directly due to its objective function, which consists of a weighted sum of logarithmic terms. 
In addition to high-complexity AO algorithms, such as~\cite{Xu2024Intelligent,Du2025Intelligent}, the GNN-based optimization method can be employed to obtain a high-quality beamforming solution. As illustrated in Fig. \ref{Fig2}, the proposed method involves: extracting input features from the considered network, designing the GNN architecture, defining the output, and performing model training and inference.\par

\subsubsection{Input and Output}\label{sssec:inout}
The neural network takes as input a graph \( \mathcal{G} = (\mathcal{V}, \mathcal{E}) \), where \( \mathcal{V} \) and \( \mathcal{E} \) represent the sets of nodes and edges, respectively. Upon analyzing the considered network, it is observed that the channels from each BS to its associated UEs are mutually dependent. Therefore, the CSI of these channels is modeled as node features within a fully connected (FC) graph. In accordance with the number of cells, the node feature matrix $\textbf{X}^\text{in}$ is constructed by concatenating $I$ sub-matrices $\textbf{X}_{i}^\text{in}$ with $\textbf{X}^\text{in} = [\textbf{X}_{1}^\text{in}, \dots, \textbf{X}_{i}^\text{in}, \dots,  \textbf{X}_{I}^\text{in}]$, where $\textbf{X}_{i}^\text{in}$ is given by
\begin{align}
\textbf{X}_i^\text{in} = \begin{bmatrix}
\text{Re} \{ \textbf{h}_{i1, i} \} & \text{Im} \{ \textbf{h}_{i1, i} \} \\
\vdots & \vdots   \\
\text{Re} \{ \textbf{h}_{ik_i, i} \} & \text{Im} \{ \textbf{h}_{ik_i, i}\} \\
\vdots & \vdots   \\
\text{Re} \{ \textbf{h}_{iK_i, i} \} & \text{Im} \{ \textbf{h}_{iK_i, i} \} \\
\end{bmatrix}.
\end{align} 
The neural network's final output is the beamforming matrix $\textbf{X}^\text{out} = [\textbf{X}_{1}^\text{out}, \dots, \textbf{X}_{i}^\text{out}, \dots,  \textbf{X}_{I}^\text{out}]$, where $\textbf{X}_{i}^\text{out}$ is given by
\begin{align}
\textbf{X}_i^\text{out} = \begin{bmatrix}
\text{Re} \{ \textbf{w}_{i1, i} \} & \text{Im} \{ \textbf{w}_{i1, i} \} \\
\vdots & \vdots   \\
\text{Re} \{ \textbf{w}_{ik_i, i} \} & \text{Im} \{ \textbf{w}_{ik_i, i}\} \\
\vdots & \vdots   \\
\text{Re} \{ \textbf{w}_{iK_i, i} \} & \text{Im} \{ \textbf{w}_{iK_i, i} \} \\
\end{bmatrix}.
\end{align} 

\subsubsection{Neural Network Architecture}\label{sssec:nna}
The overall architecture integrates multiple homogeneous GNNs, where each GNN is responsible for a specific cell and receives as input the CSI from that cell's BS to its associated UEs. This multi-GNN structure is ideal for distributed deployment, allowing each GNN to operate locally at its corresponding BS, which can significantly reduce inference latency by avoiding centralized processing. Since all the GNNs are homogeneous---sharing the same architecture and operational rules, differing only in model parameters---we proceed to describe the structure and operation of a single GNN.

A single GNN consists of a multi-layer perceptron (MLP) as the input layer, followed by two identical successive GNN layers, and ending with a FC layer for the output. The initial MLP layer takes in $2N$ input neurons ($N$ being the number of FAs in a cell), corresponding to the number of columns in the node feature sub-matrix. The output of the FC layer aligns with the dimensionality of the beamforming sub-matrix, resulting in $2N$ output neurons. Between the initial MLP and final FC layers lie two GNN layers, serving as the core of the neural network, which extract and propagate essential graph features through aggregation and combination mechanisms.

\subsubsection{Operational Rule of GNN Layers}\label{sssec:operate}
After the input data $\textbf{X}_i^\text{in}$ passes through the initial MLP layer, the resulting output $\textbf{X}_i^{(1)}$ is fed into the first GNN layer. The GNN layer contains two MLP sub-layers coupled with aggregation and combination operations. The operational procedure is implemented as follows:
\begin{itemize}
\item [\ding{172}] For any given row \( k \in K \) of the feature matrix \( \textbf{X}_i^{(1)} \), the remaining rows are passed through the first MLP in the GNN layer, which is mathematically is given by
\begin{equation}
\textbf{x}_{i,k'}^{(2)} = \text{MLP1}\left( \textbf{x}_{i,k'}^{(1)}\right), ~ {k' \in \mathcal{K} \setminus k}.
\end{equation}
where $\textbf{x}_{i,k'}^{(1)}$ denotes the $k'$-th row vector of \( \textbf{X}_i^{(1)} \) and $\textbf{x}_{i,k'}^{(2)}$ represents the corresponding output. 
\item [\ding{173}] Element-wise max-pooling along columns is performed to the matrix formed by concatenating all \( \textbf{x}_{i,k'}^{(2)} \)  vectors, which is given by
\begin{equation}
\textbf{x}_{i,k'}^\text{max} = \text{MaxPool}\left( \textbf{x}_{i,k'}^{(2)}\right), ~ {k' \in \mathcal{K} \setminus k}.
\end{equation}
Through \ding{172} and \ding{173}, the aggregation is performed.
\item [\ding{174}] Subsequently, $\textbf{x}_{i,k}^{(1)}$ and $\textbf{x}_{i,k'}^\text{max}$ are concatenated along the row direction, which is given by
\begin{equation}
\textbf{x}_{i,k}^{(3)} = [\textbf{x}_{i,k}^{(1)}, \textbf{x}_{i,k'}^\text{max}], ~ {k \in \mathcal{K} } ~ \text{and} ~ {k' \in \mathcal{K} \setminus k}.
\end{equation}
 \item [\ding{175}] The feature representation $\textbf{x}_{i,k}^{(3)}$ is processed through the second MLP sub-layer to generate the output of GNN layer $\textbf{x}_{i,k}^{(4)}$, which is mathematically is given by
\begin{equation}
\textbf{x}_{i,k}^{(4)} = \text{MLP2}\left( \textbf{x}_{i,k}^{(3)} \right), ~ {k \in \mathcal{K}}.
\end{equation}
Through \ding{174} and \ding{175}, the combination is performed and the output feature matrix \( \textbf{X}_i^{(4)} \) is obtained with $\textbf{x}_{i,k}^{(4)}$.)
 \end{itemize} 
The operations of the second GNN layer are omitted here to avoid redundancy, as they mirror those of the first GNN layer.

Generally, the aggregation and combination operations critically determine GNN's scalability. In the dual GNN layers, a composite aggregation function (MLP + max-pooling) is used to extract and propagate neighborhood features, capturing UE's interactions. The combination function concatenates features followed by MLP transformation. Through such a design, the model learns UE's interaction patterns for interference suppression. Notably, the architecture generalizes to varying UE counts by concatenating all channels into a unified input tensor, ensuring configuration-independent operation.

\subsubsection{Centralized Training and Distributed Deployment}\label{sssec:train}
For the FA-enabled downlink MIMO multi-cell scenario, the neural network consists of $I$ parallel GNNs, where each GNN is dedicated to a specific BS. The training process is conducted centrally to optimize model parameters, after which each trained GNN module is independently deployed to its respective BS for distributed inference operations.

During model training, the integrated multi-GNN architecture undergoes centralized optimization, allowing full access to global CSI. The neural network's weight and bias parameters are adjusted through unsupervised learning, with parameter updates guided by a loss function defined as
\begin{align}
\mathcal{L} =  - \frac{\sum_{t \in \mathcal{T}}   \sum_{i \in \mathcal{I}, k \in \mathcal{K}_i}  \omega_{ik} R_{ik}^{(t)}}{T}, \nonumber 
\end{align} 
where $R_{ik}^{(t)}$ is the achievable communication rate at the $k$-th UE in the $i$-th cell for the $t$-th channel sample, and $T$ denotes the cardinality of the training set $ \mathcal{T}$, representing the total number of available samples for model optimization. The formulated objective function in (P2) exhibits an inverse relationship with the loss function. Through iterative stochastic gradient descent (SGD) optimization, minimization of the loss function corresponds to asymptotic maximization of the objective function. The multi-GNN architecture, leveraging global CSI availability, effectively learns to characterize both inter- and intra-cell interference patterns. The offline nature of this training phase alleviates practical computational constraints.

During model inference, the multi-GNN architecture operates in a semi-distributed manner. The global model is partitioned into multiple GNN instances, with each BS equipped with its dedicated GNN accelerator or processor. This design enables distributed computation of beamforming vectors at each BS using only local CSI and its respective GNN module. With global CSI availability, BSs can engage in coordinated beamforming, enabling network-wide performance optimization through distributed cooperation. This collaborative approach maximizes the overall system capacity while maintaining individual BS autonomy. In short, semi-distributed implementation refers to a hybrid architecture in which GNN inference is executed distributively at each BS while inter-BS coordination is performed centrally.

\subsubsection{Computational Complexity}\label{sssec:complexity}

\begin{algorithm}[t]
\caption{Implementation of A Single GNN}\label{alg1}
\begin{algorithmic}[1]
\State \textbf{Input}: $\textbf{X}_i^\text{in}$ for $i \in \mathcal{I}$
\State \textbf{Output}: $\textbf{X}_{i}^\text{out}$ for $i \in \mathcal{I}$
\State $\textbf{X}_{i}^{(1)} \gets \text{MLP}^\text{in}\left( \textbf{X}_i^\text{in} \right)$

\For{$k = 1$ to $K_i$}
    \For{$k' = 1$ to $K_i$}
        \If{$k' \ne k$}
            \State $\textbf{x}_{i,k'}^{(2)}  \gets \text{MLP1}\left( \textbf{x}_{i,k'}^{(1)}\right)$
        \EndIf
    \EndFor
    \State $\textbf{x}_{i,k'}^\text{max} \gets \text{MaxPool}\left( \textbf{x}_{i,k'}^{(2)}\right),  ~ {k' \in \mathcal{K} \setminus k}.$
    \State $\textbf{x}_{i,k}^{(3)} \gets [\textbf{x}_{i,k}^{(1)}, \textbf{x}_{i,k'}^\text{max}]$
    \State $\textbf{x}_{i,k}^{(4)} \gets \text{MLP2}\left( \textbf{x}_{i,k}^{(3)} \right)$
\EndFor

\For{$k = 1$ to $K_i$}
    \For{$k' = 1$ to $K_i$}
        \If{$k' \ne k$}
            \State $\textbf{x}_{i,k'}^{(5)}  \gets \text{MLP3}\left( \textbf{x}_{i,k'}^{(4)}\right)$.
        \EndIf
    \EndFor
    \State $\textbf{x}_{i,k'}^\text{max} \gets \text{MaxPool}\left( \textbf{x}_{i,k'}^{(5)}\right), ~ {k' \in \mathcal{K} \setminus k}$
    \State $\textbf{x}_{i,k}^{(6)} \gets [\textbf{x}_{i,k}^{(4)}, \textbf{x}_{i,k'}^\text{max}]$
    \State $\textbf{x}_{i,k}^{(7)} \gets \text{MLP4}\left( \textbf{x}_{i,k}^{(6)} \right)$
\EndFor

\State $\textbf{X}_{i}^{(8)} \gets \text{FC} \left(  \textbf{X}_{i}^{(7)}  \right)$
\State $ \textbf{X}_{i}^\text{out} \gets \sqrt{P} \text{LayerNorm} \left( \textbf{X}_{i}^{(8)}  \right)$
\end{algorithmic}
\end{algorithm}

The implementation details of the adopted GNN instance are illustrated in Algorithm \ref{alg1}. The inference latency of the multi-GNN model critically impacts the real-time performance of FA-assisted multi-cell MIMO downlink network, necessitating rigorous computational complexity analysis of its forward pass. By virtue of architectural homogeneity across all constituent GNNs, the computational analysis of any single GNN module sufficiently characterizes the entire ensemble without loss of generality. Taking the $i$-th GNN module as an example for illustration, the input processing stage employs an MLP layer. The resulting computational complexity for this component is $\mathcal{O}(2K_i NL_1 + K_i L_1L_2)$, where $L_1$ indicates the dimensionality of the hidden representation and $L_2$ specifies the output dimension. Each processing unit in the MLP incorporates a nonlinear activation function. The architecture employs two GNN layers, each containing two MLP layers along with their respective activation functions. Taking the aggregation and combination operations into account, the combined computational complexity for these layers is approximately given by $\mathcal{O}\big(2K_i \big[(K_i-1)K_i (L_3L_4 + L_4L_5) + K_i(L_6L_7 + L_7L_8)\big]\big)$, where $L_3$, $L_4$, and $L_5$ indicate the neuron counts in the input, hidden, and output layers of the first MLP, while $L_6$, $L_7$, and $L_8$ correspond to those in the second MLP. The output layer of each GNN is realized through an FC layer, contributing a computational complexity of $\mathcal{O}(2K_iL_8N)$. The overall computational complexity is essentially the summation of the complexities from these constituent parts.

\subsection{RPS}\label{ssec:random}
As previously formulated, the optimization problem (P1) can be decomposed into two subproblems: beamforming optimization and port selection. The preceding subsection has presented the GNN-based solution for the beamforming optimization. This subsection presents the integration of RPS and GNN-based beamforming.

In the RPS strategy, a single port is randomly selected from each FA's available port set. The resulting combination of randomly chosen set of ports is subsequently employed for GNN-based beamforming optimization. In order to optimize performance, this random selection process is repeated across multiple independent trials, with each trial's outcome being evaluated to ultimately identify the highest-performing port selection. In the simulation section (see Section \ref{sec:evaluate} for details), it can be observed that the WSR distribution with varying numbers of selections demonstrates that RPS, when independent trials are conducted tens of times, can achieve a suboptimal performance very close to the optimal solution.

\section{Instruction-Driven FPGA-based Acceleration}\label{sec:fpga}
To minimize the inference latency, in this section, we design a specialized FPGA-based accelerator featuring a domain-specific ISA and an optimized overlay micro-architecture. In addition, an efficient scheduling algorithm is also developed to speed up concurrent GNN inference tasks across multiple port selections on the FPGA platform. 

\begin{figure*}
\centering
\includegraphics[width=6.5in]{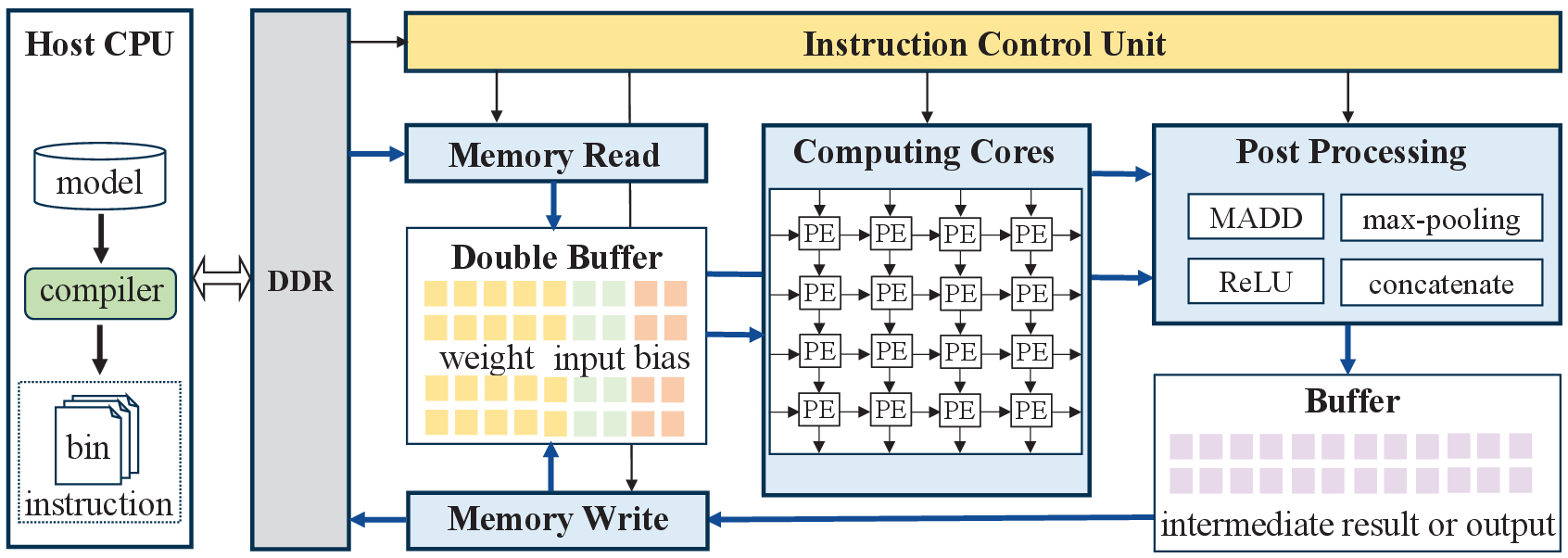}
\caption{An illustration of customized ISA.}\label{Fig3}
\end{figure*}

\subsection{Customized ISA}\label{ssec:isa}
To reduce latency, improve energy efficiency, and enhance control flexibility, the FPGA-based GNN accelerator adopts a complex instruction set computer (CISC)-style ISA, which supports three types of instructions.
\begin{itemize}
\item \texttt{Memory Access}: This instruction type manages data movement between off-chip and on-chip memory, with read and write operations determined by the memory-type field of the instruction, and the corresponding memory addresses explicitly specified.
\item \texttt{Matrix Processing}: This instruction type drives multiple systolic arrays (SAs) to efficiently execute matrix multiplication, which is a fundamental operation in GNN inference. By optimizing data flow and enabling parallel processing, it accelerates large-scale matrix multiplications on hardware.
\item \texttt{Post Processing}: This instruction type handles operations beyond matrix multiplication, including matrix addition (MADD), activation functions, max-pooling, concatenation, and normalization, in order to prepare the output for subsequent computation or storage.
\end{itemize}

\subsection{Micro-architecture}\label{ssec:micro}
Based on the aforesaid ISA, an overlay micro-architecture is designed, see Fig.~\ref{Fig3}, consisting of four key units: control, memory read/write, matrix processing, and post processing.

\subsubsection{Control Unit}\label{sssec:control}
The control unit is responsible for fetching and prefetching instructions from off-chip memory, decoding the instruction types, and dispatching them to the corresponding functional units, including memory read, matrix processing, post processing, and memory write, to coordinate the overall execution process. To manage both host interaction and on-chip execution efficiently, the control module is architecturally divided into external and internal control logic. The external control is responsible for handling host-side commands, configuration, and global execution control, such as triggering the accelerator and managing data transfers through interfaces like PCIe or AXI. In contrast, the internal control logic operates within the accelerator cores, decoding custom instructions, coordinating the operation of memory, matrix processing, and post processing units, and managing data dependencies. This separation improves design modularity, simplifies system integration across different platforms, and enables flexible adaptation to diverse GNN workloads by decoupling high-level control from low-level execution management.

\subsubsection{Memory Read/Write Units}\label{sssec:memory}
The memory read/write units manage data transfers between off-chip memory (e.g., DRAM) and on-chip buffers (e.g., BRAM or SRAM), serving as a bridge between external data and internal compute modules. Tiled inputs and model parameters are transferred from off-chip memory to the on-chip local buffers via AXI-stream to ensure efficient data access. To achieve parallelism between data loading and computation, a ping-pong buffering mechanism is adopted. Two alternating buffers allow one buffer to load data while the other feeds the compute units, enabling concurrent data movement and computation. This approach ensures continuous streaming of inputs and outputs, beneficial in GNN inference.

\subsubsection{Matrix Processing Unit}\label{sssec:mpu}
In the adopted GNN, each layer fundamentally relies on FC operations, which are inherently dependent on large-scale matrix multiplications. To efficiently execute these computations, the FPGA-based GNN accelerator employs multiple SAs as its core computation unit. An SA is a specialized hardware architecture consisting of a grid of processing elements (PEs). Each PE is responsible for computing partial products, accumulating intermediate results, and rhythmically passing data to adjacent elements. This synchronized, pulse-like data flow enables highly parallel processing, making SAs particularly well-suited for workloads that involve intensive and repetitive matrix operations.

\begin{figure*}
\centering
\includegraphics[width=7in]{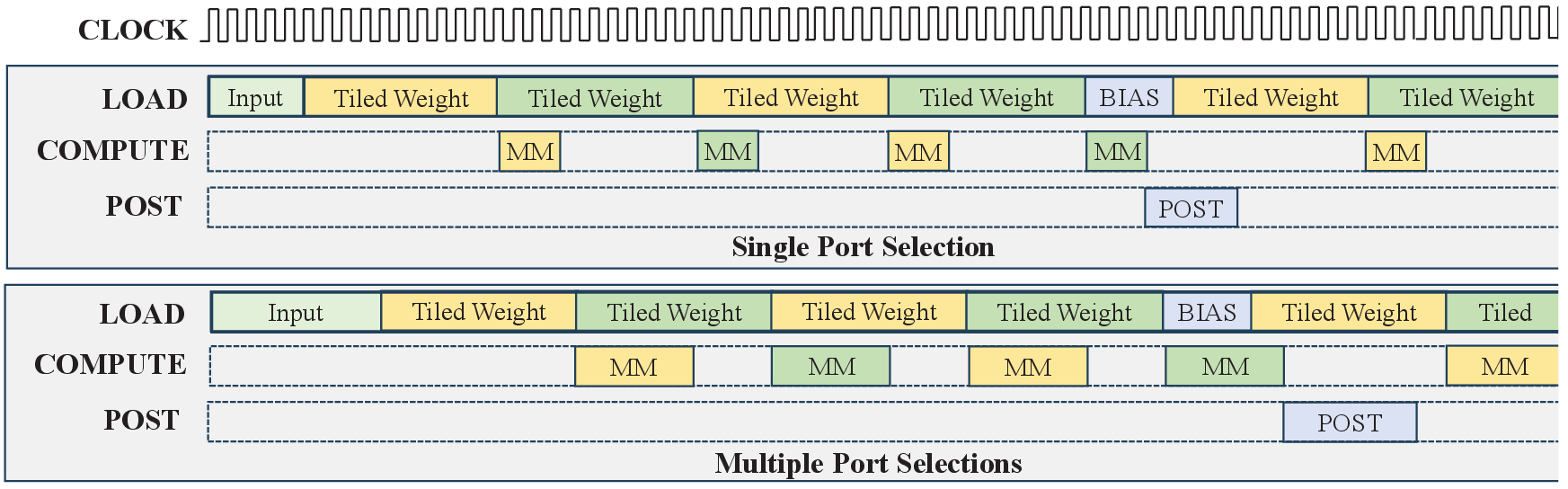}
\caption{An illustration of hardware micro-architecture.}\label{Fig4}
\end{figure*}

\subsubsection{Post Processing Unit}\label{sssec:ppu}
The post processing unit includes operations such as MADD, activation function, max-pooling, concatenation, normalization, and other related functions. Once data passes through this module, the output is managed according to its purpose: intermediate results are stored in a dedicated buffer and subsequently fed back into the computing cores, while final results are written back to external memory. This design fundamentally improves dataflow efficiency by reducing reliance on external memory, enabled by a clear separation between intermediate and final outputs. By minimizing intermediate data transfers between on-chip and off-chip memory, it significantly lowers latency and reduces the energy overhead caused by excessive data movement. 

\begin{algorithm}[t]
\caption{Code Refactoring for A Single GNN}\label{alg2}
\begin{algorithmic}[1]
\State \textbf{Input}: $\textbf{X}_i^\text{in} (\textbf{s1})$, $\textbf{X}_i^\text{in} (\textbf{s2})$ for $i \in \mathcal{I}$, where $\textbf{s1}$, $\textbf{s2}$ $\in \mathcal{S}$
\State \textbf{Output}: $\textbf{X}_{i}^\text{out}$  for $i \in \mathcal{I}$
\State $\textbf{X}_{i}^\text{in} \gets \text{concatenate} \left( \textbf{X}_i^\text{in}  (\textbf{s1}); \textbf{X}_i^\text{in}  (\textbf{s2})  \right)$

\State $\textbf{X}_{i}^{(1)} \gets \text{MLP}^\text{in} \left( \textbf{X}_i^\text{in}  \right)$
\State $\textbf{X}_{i}^{(2)}  \gets \text{MLP1}\left(\textbf{X}_{i}^{(1)} \right)$

\For{$k = 1$ to $K_i$}
    \State $\textbf{x}_{i,k'}^\text{max} (\textbf{s1}) \gets \text{MaxPool}\left( \textbf{x}_{i,k'}^{(2) }  (\textbf{s1}) \right),  ~ {k' \in \mathcal{K} \setminus k}.$
    \State $\textbf{x}_{i,k}^{(3)} (\textbf{s1}) \gets [\textbf{x}_{i,k}^{(1) }  (\textbf{s1}), \textbf{x}_{i,k'}^\text{max} (\textbf{s1})]$
    \State $\textbf{x}_{i,k'}^\text{max} (\textbf{s2}) \gets \text{MaxPool}\left( \textbf{x}_{i,k'}^{(2) }  (\textbf{s2}) \right),  ~ {k' \in \mathcal{K} \setminus k}.$
    \State $\textbf{x}_{i,k}^{(3)} (\textbf{s2}) \gets [\textbf{x}_{i,k}^{(1) }  (\textbf{s2}), \textbf{x}_{i,k'}^\text{max} (\textbf{s2})]$
\EndFor

\State $\textbf{X}_{i}^{(3)} \gets \text{concatenate} \left( \textbf{X}_i^{(3)}  (\textbf{s1}); \textbf{X}_i^{(3)}  (\textbf{s2})  \right)$
\State $\textbf{X}_{i}^{(4)}  \gets \text{MLP2}\left( \textbf{X}_{i}^{(3)} \right)$

\State $\textbf{X}_{i}^{(5)}  \gets \text{MLP3}\left(\textbf{X}_{i}^{(4)} \right)$
\For{$k = 1$ to $K_i$}
    \State $\textbf{x}_{i,k'}^\text{max} (\textbf{s1}) \gets \text{MaxPool}\left( \textbf{x}_{i,k'}^{(5) }  (\textbf{s1}) \right),  ~ {k' \in \mathcal{K} \setminus k}.$
    \State $\textbf{x}_{i,k}^{(6)} (\textbf{s1}) \gets [\textbf{x}_{i,k}^{(4) }  (\textbf{s1}), \textbf{x}_{i,k'}^\text{max} (\textbf{s1})]$
    \State $\textbf{x}_{i,k'}^\text{max} (\textbf{s2}) \gets \text{MaxPool}\left( \textbf{x}_{i,k'}^{(5) }  (\textbf{s2}) \right),  ~ {k' \in \mathcal{K} \setminus k}.$
    \State $\textbf{x}_{i,k}^{(6)} (\textbf{s2}) \gets [\textbf{x}_{i,k}^{(4) }  (\textbf{s2}), \textbf{x}_{i,k'}^\text{max} (\textbf{s2})]$
\EndFor

\State $\textbf{X}_{i}^{(6)} \gets \text{concatenate} \left( \textbf{X}_i^{(6)}  (\textbf{s1}); \textbf{X}_i^{(6)}  (\textbf{s2})  \right)$
\State $\textbf{X}_{i}^{(7)} \gets \text{MLP4}\left( \textbf{X}_{i}^{(6)} \right)$

\State $\textbf{X}_{i}^{(8)} \gets \text{FC} \left( \textbf{X}_{i}^{(7)} \right)$
\State $ \textbf{X}_{i}^\text{out} \gets \sqrt{P} \text{LayerNorm} \left( \textbf{X}_{i}^{(8)}\right)$
\end{algorithmic}
\end{algorithm}

\subsection{Task Scheduling Algorithm}\label{ssec:task}
According to the roofline model~\cite{Zhang2015Optimizing}, the primary factors limiting inference latency in FPGA-based accelerators are computational throughput and the bandwidth constraints between on-chip and off-chip memory. Given the substantial number of weight and bias parameters in the considered GNN, the inference latency of the FPGA-based accelerator is predominantly determined by the I/O bandwidth for accessing off-chip memory. This makes the system memory-bound rather than computation-bound. This issue is particularly critical to the performance of the proposed FPGA-based accelerator, as the substantial memory demands of GNNs often lead to data access becoming the dominant bottleneck, thereby overshadowing the computational workload.

In addition to frequently used optimization techniques such as quantization, loop tiling and double buffering, efficient task scheduling plays a crucial role in improving computational efficiency by enabling the concurrent inference of multiple GNN tasks. As shown in Fig.~\ref{Fig4}, GNN is a memory-intensive task and the computing cores often remain idle, waiting for data from off-chip memory. In the considered communication scenario, each FA port selection corresponds to a separate GNN inference required to compute the beamforming solution. As the number of port selections increases, performing these inference tasks sequentially leads to a linear growth in latency. For different port selections, the corresponding GNN inference tasks can share the same model and identical parameters. This enables concurrent parallel inference of these GNN tasks, which in turn maximizes overall throughput and resource utilization. This is because off-chip memory accesses are significantly reduced through parameter sharing, and redundant computations are eliminated. More importantly, the computing cores are prevented from stalling due to data transfer latency.

As shown in Algorithm \ref{alg2}, the computation flow of multiple GNN inference tasks on the FPGA-based accelerator is illustrated using two tasks as an example. First, the GNN model parameters, channel data for multiple port selections, and the instruction stream are transferred from off-chip memory to on-chip buffers. Second, the channel data is concatenated on-chip. Third, these inputs are processed by the first MLP layer of the GNN, where matrix multiplication and post processing are performed. The results are stored in an intermediate buffer. Fourth, the intermediate results are used by the two GNN layers. Through code restructuring and reuse of intermediate computations, compared with Algorithm \ref{alg1}, this stage achieves higher efficiency. Finally, the outputs from the GNN layers are passed through an FC layer and a normalization layer to generate the final inference results, which are then written back to off-chip DDR memory. Throughout the entire process, all computations are carried out on-chip, effectively avoiding repeated off-chip memory access and reducing overall latency.

\section{Evaluation}\label{sec:evaluate}
This section provides a comprehensive evaluation of the communication and computational performance of the proposed hardware, software co-design, developed for beamforming and port selection in the FA-enhanced multi-cell MIMO network. The communication performance of the GNN-RPS approach is first examined through numerical simulations. Subsequently,  the computational performance of the designed FPGA-based GNN accelerator, which incorporates an efficient scheduling algorithm, is evaluated through experiments.

\subsection{Simulation and GNN Configuration}\label{ssec:simulate}

\begin{table}[t]
\scriptsize
\begin{center}
\caption{Simulation Parameters.}\label{T1}~~\\
\begin{tabular}{c|c|c} 
\hline
\textbf{Notation}        &  \textbf{Description}              & \textbf{Value}      \\ \hline\hline
$I$                      &  Number of cells                   & $2$                   \\ \hline
$K$                      &  Number of UEs per cell            & $4$                   \\ \hline
$r$                      &  Positions  of all the UEs         & $[20~{\rm m}, 30~{\rm m}]$        \\ \hline
$d_0$                    &  Reference distance                & $1~{\rm m}$                 \\ \hline
$N$                      &  Number of FAs at each BS          & $4$                   \\ \hline
$L$                      &  Number of ports at each FA        & $6$                   \\ \hline
$W \lambda $             &  Length of each FA                 & $0.5 \lambda $      \\ \hline
$P$                      &  Transmit power at each BS         & $3~{\rm dBm}$               \\ \hline
$\sigma^2$               &  Noise variance at UEs             & $-90~{\rm dBm}$             \\ \hline
$\omega_{m}$             &  Weighting factor                  & $1$                   \\ \hline
\end{tabular}
\end{center}
\end{table}

The simulation parameters for the FA-enhanced downlink multi-cell MIMO network are summarized in Table \ref{T1}. Specifically, the network consists of \(I = 2\) cells, each serving an equal number of UEs, with \(K = 4\) UEs per cell. Each BS is equipped with \(N = 4\) FAs, each comprising \(L = 6\) ports. The available ports in each FA are assumed to be evenly distributed along a line segment of length \(W \lambda = 0.5 \lambda\). The positions $r$ of all UEs are randomly generated within the range of $20~{\rm m}$ to $30~{\rm m}$. Let \(d_x\) and \(d_0\) denote the transmission distance and the reference distance, respectively. The path loss is then given by \( \delta = \delta_0 - 25 \log_{10}\left( \frac{d_x}{d_0} \right) \, \text{dB}, \) where \(\delta_0 = -30 \, \text{dB}\) represents the path loss at the reference distance. The transmit power is set to \(P = 3 \, \text{dBm}\), and the noise variance at each UE is set to \(\sigma^2 = -90 \, \text{dBm}\).  The weighting factors are defined as \(\omega_m = 1\). Some parameters can be varied within a range when used as the $x$-axis in the simulations.

\begin{table*}[t]
\scriptsize
\begin{center}
\caption{GNN Setup.}\label{TII}~~\\
\begin{tabular}{cc||cccc||cccc||c}
\hline
\multicolumn{2}{c||}{\textbf{MLP}}                              & \multicolumn{4}{c||}{\textbf{GNN layer} }                                                                  & \multicolumn{4}{c||}{\textbf{GNN layer} }                                                                    & \multirow{3}{*}{\textbf{FC}} \\ \cline{1-10}
\multicolumn{1}{c|}{\multirow{2}{*}{FC}} & \multirow{2}{*}{FC} & \multicolumn{2}{c|}{MLP}                                    & \multicolumn{2}{c||}{MLP}                & \multicolumn{2}{c|}{MLP}                                    & \multicolumn{2}{c||}{MLP}                &                              \\ \cline{3-10}
\multicolumn{1}{c|}{}                    &                     & \multicolumn{1}{c|}{FC}      & \multicolumn{1}{c|}{FC}      & \multicolumn{1}{c|}{FC}       & FC      & \multicolumn{1}{c|}{FC}      & \multicolumn{1}{c|}{FC}      & \multicolumn{1}{c|}{FC}       & FC      &                              \\ \hline
\multicolumn{1}{c|}{$2N \times 1024$}             & $1024 \times 512$            & \multicolumn{1}{c|}{$512 \times 512$} & \multicolumn{1}{c|}{$512 \times 512$} & \multicolumn{1}{c|}{$1024 \times 512$} & $512 \times 512$ & \multicolumn{1}{c|}{$512 \times 512$} & \multicolumn{1}{c|}{$512 \times 512$} & \multicolumn{1}{c|}{$1024 \times 512$} & $512 \times 512$ & $512 \times 2N$                 \\ \hline
\end{tabular}
\end{center}
\end{table*}

In the adopted multi-GNN architecture, all GNN instances share an identical structural design. Each GNN comprises one initial MLP, two GNN layers, and one FC layer, with the detailed hyperparameter settings presented in Table \ref{TII}. ReLU activation functions are applied to all FC layers throughout the network, except for the final output layer. The model is trained using the Adam optimizer, initialized with a learning rate of $0.001$. A learning rate scheduling strategy is employed, where the rate is decayed by a factor of $0.995$ every $100$ training steps. Each training epoch consists of $10,000$ randomly generated samples, which are divided into batches of $200$, resulting in $50$ updates per epoch. The test set contains $2,000$ samples. Training terminates either when a predefined number of epochs is reached or when convergence criteria are met.

\subsection{Communication Performance}\label{ssec:perform}
The achievable communication performance of the proposed GNN-RPS approach is evaluated through simulations and compared with several benchmark methods. The details of all considered schemes are outlined below.

\begin{itemize}
\item \texttt{GNN-RandomMax}: This is the proposed GNN-RPS in which RPS is evaluated over $20$ independent trials, with the best-performing configuration retained.
\item \texttt{GNN-RandomSingle}: This is the proposed GNN-RPS in which RPS is executed once with a one-time trial, resulting in a fixed antenna configuration equivalent.
\item \texttt{GNN-Exhaustive}: This is the proposed GNN-RPS but $500$ independent trials are conducted to approximate an exhaustive search for a performance upper bound.
\item \texttt{MMSE-Exhaustive}: The conventional MMSE-based approach is combined with $500$ independent port selections, serving as a non-learning benchmark.
\item \texttt{MRT-Exhaustive}: The MRT-based scheme is combined with $500$ independent port selections.
\item \texttt{ZF-Exhaustive}: The ZF-based transmission method is integrated with $500$ independent port selections.
\end{itemize}

\begin{figure}[t]
\centering
\includegraphics[width=\linewidth]{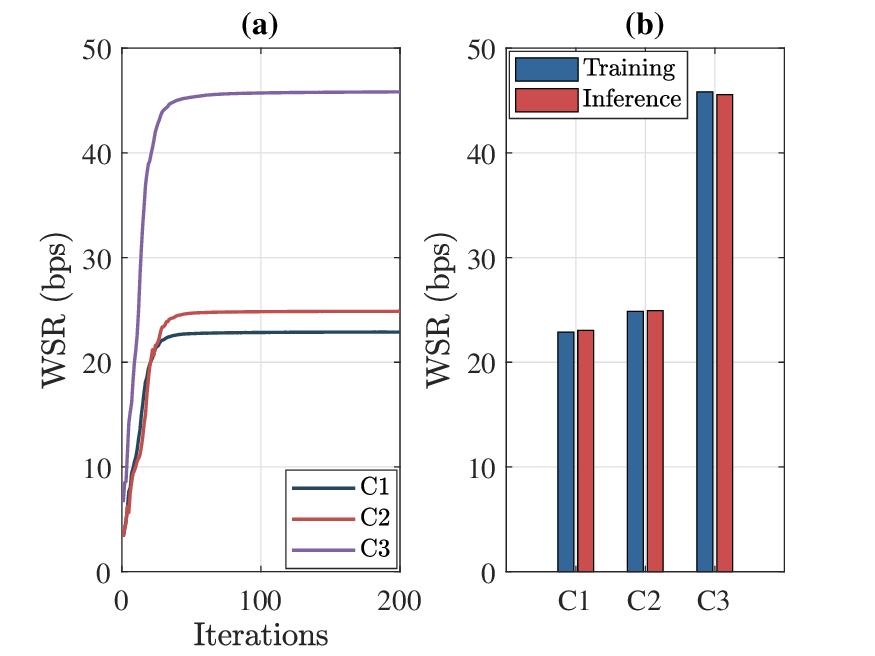}
\caption{Training and inference performance of the multi-GNN method under different parameter settings $(I, K, P)$: $C_1 = (2,2,0)$, $C_2 = (2,2,3)$, and $C_3 = (4,2,0)$.}\label{Fig5}
\end{figure}

In Fig.~\ref{Fig5}, we illustrate the training and inference performance of the multi-GNN method under different parameter settings $(I, K, P)$: $C_1 = (2,2,0)$, $C_2 = (2,2,3)$, and $C_3 = (4,2,0)$. The results indicate that the GNN converges rapidly, typically stabilizing within $30$ iterations, with only a slight improvement observed thereafter. To evaluate the model's generalization capability, the training and inference results are compared. The close agreement between these results demonstrates the robustness and consistent performance of the model across various channel conditions.

\begin{figure*}
\centering
\includegraphics[width=\linewidth]{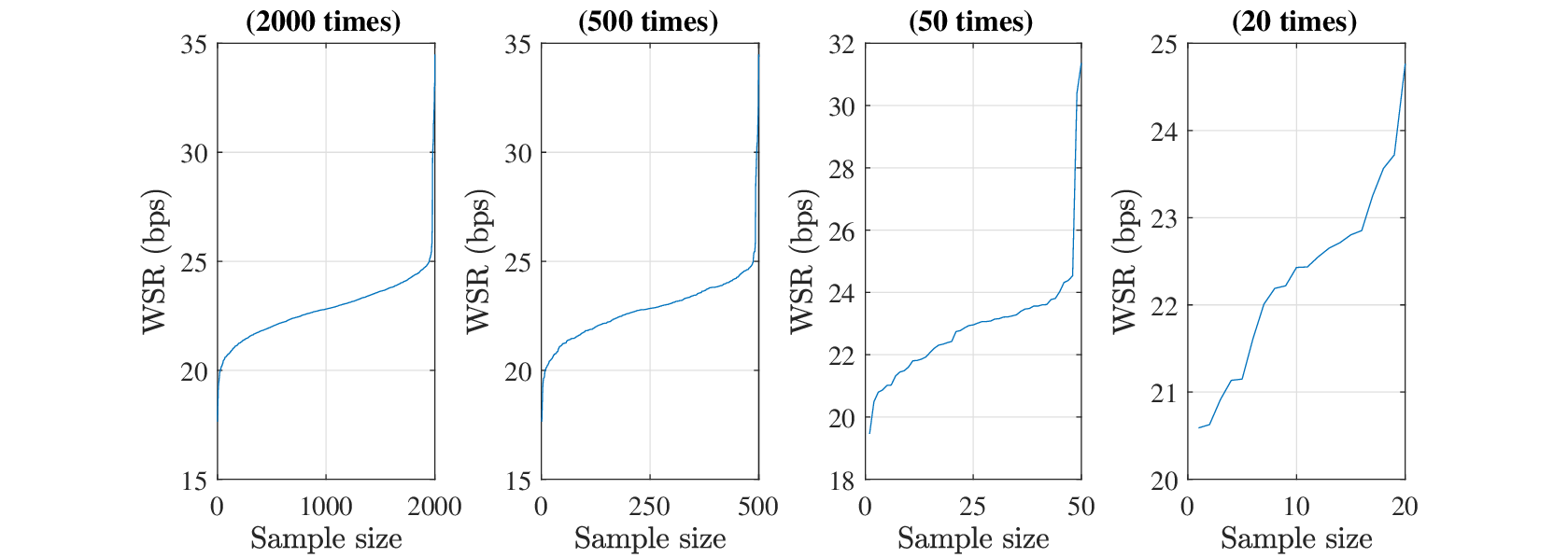}
\caption{WSR distribution with varying numbers of RPSs.}\label{Fig6}
\end{figure*}

Fig.~\ref{Fig6} presents the WSR distribution under different numbers of RPS trials. As observed from the figure, the maximum achievable WSR generally increases with the number of sampling trials, indicating that more extensive exploration tends to yield better performance. However, the performance gain becomes marginal when the number of samples increases from $500$ to $2000$, suggesting diminishing returns. Notably, even with as few as $20$ trials, the WSR reaches approximately $70\%$ of the value achieved with $2000$ trials. This implies that a relatively small number of samples can still provide a near-optimal solution with significantly reduced computational cost. Therefore, a clear trade-off exists between performance and computational complexity in the port selection process.

\begin{figure}
\centering
\includegraphics[width=\linewidth]{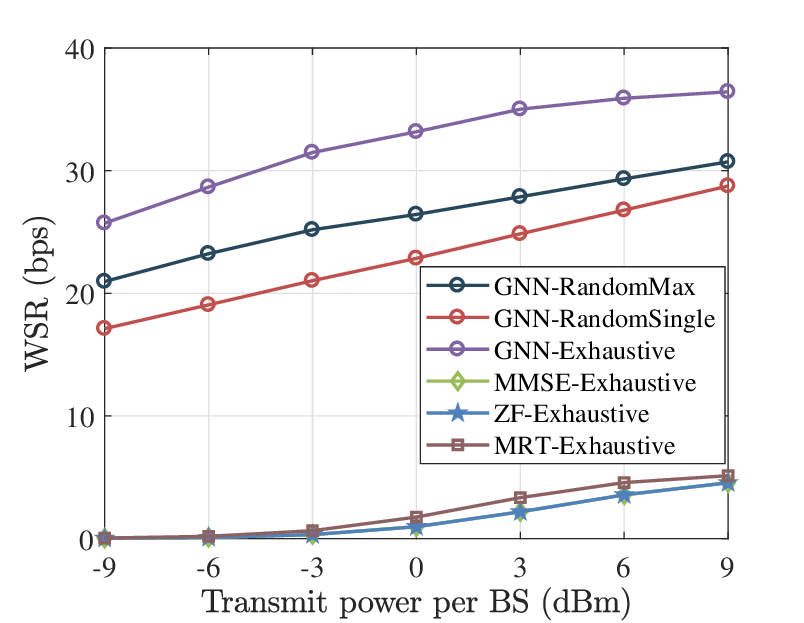}
\caption{The relationship between the WSR and the transmit power.}\label{Fig7}
\end{figure}

Fig.~\ref{Fig7} shows the relationship between the WSR and transmit power, indicating that higher power significantly improves performance. Also, the GNN-based methods outperform traditional MMSE, ZF, MRT approaches, benefiting from the multi-GNN's ability to capture inter-cell and inter-user dependencies and manage interference. Among them, \texttt{GNN-RandomMax} performs better than \texttt{GNN-RandomSingle} but is inferior to \texttt{GNN-Exhaustive} due to fewer port selection samples. This highlights the advantage of FAs in providing additional DoF through port selection. 

\begin{figure}
\centering
\includegraphics[width=\linewidth]{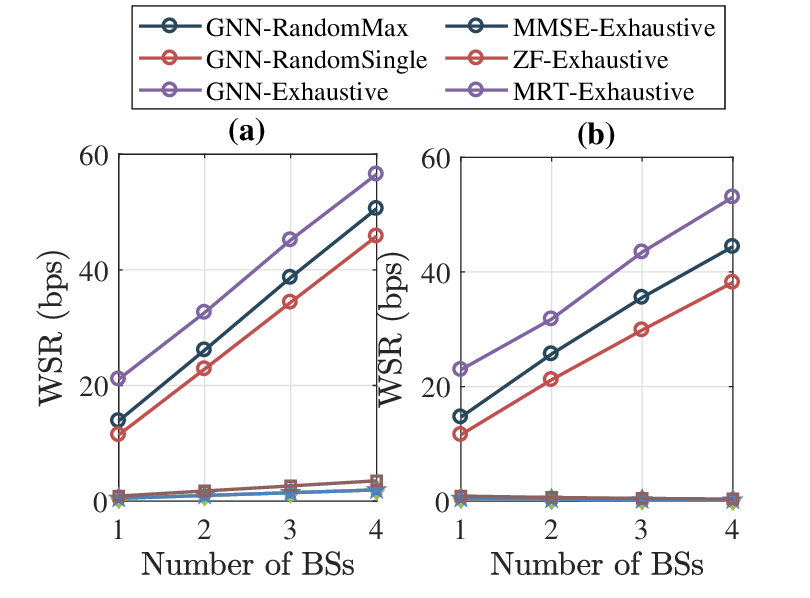}
\caption{The relationship between the WSR and the number of BSs under two cases: (a) constant transmit power per BS; (b) fixed total transmit power shared among all BSs.}\label{Fig8}
\end{figure}

In Fig.~\ref{Fig8}, the results are provided to illustrate the relationship between the WSR and the number of BSs under two cases: (a) constant transmit power per BS, and (b) fixed total transmit power for all the BSs. These two cases are considered to isolate the impact of increasing the number of BSs from the potential linear increase in total transmit power. It is observed that the WSR improves with the number of BSs in both cases, primarily due to the increased spatial DoF enabled by additional BSs. Moreover, this result further verifies the relative performance of different schemes.

\begin{figure}
\centering
\includegraphics[width=\linewidth]{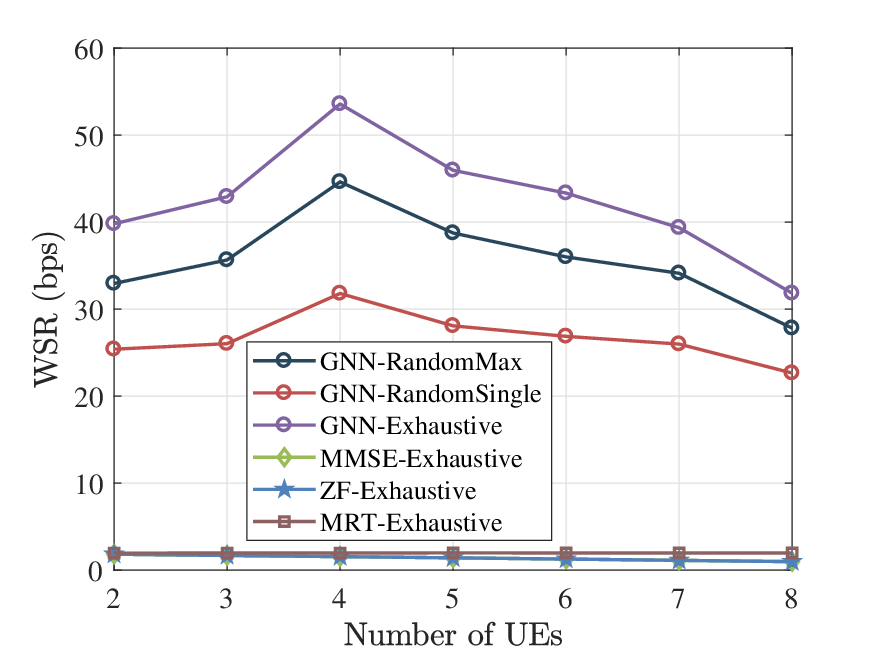}
\caption{The relationship between the WSR and the number of UEs.}\label{Fig9}
\end{figure}

Fig.~\ref{Fig9} depicts the relationship between the WSR and the number of UEs. For the GNN-based schemes, the WSR initially increases and then decreases as the number of UEs increases. The initial gain is attributed to the additional spatial DoF by having more UEs, while the performance decline results from increased intra- and inter-cell interference. This trend indicates that GNN models can effectively manage interference in low-UE-density scenarios, but their capability diminishes as the UE density increases. In contrast, traditional schemes exhibit relatively stable performance across different numbers of UEs; however, their WSR remains significantly lower than that of the GNN-based methods. This highlights the limited capability of conventional techniques including MMSE, ZF, and MRT in effectively handling interference.

\begin{figure}[]
\centering
\includegraphics[width=\linewidth]{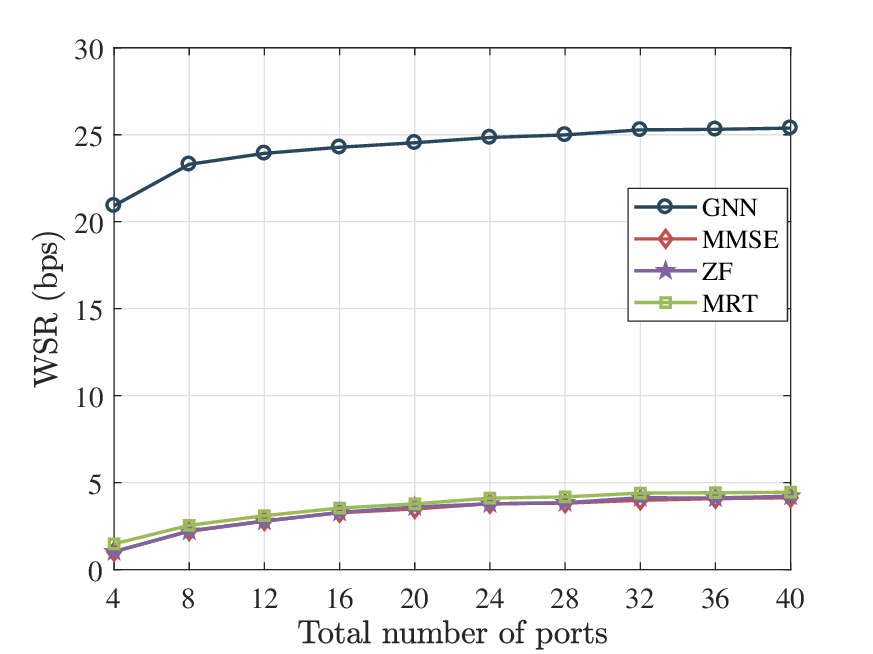}
\caption{The relationship between the WSR and the total number of ports.}\label{Fig10}
\end{figure}

Finally, Fig.~\ref{Fig10} investigates the impact of the total number of ports on the WSR, where the number of UEs per cell is $K= 2$ and each BS has $N = 2$ FAs. An exhaustive search over all possible port combinations is performed to obtain the performance upper bound. We see that the WSR increases consistently with the total number of ports. A clear performance gain is observed at the beginning, but the growth becomes slower with further increases in the number of ports. This result indicates that the spatial DoFs by port selection effectively enhance the communication performance but the marginal gains diminish noticeably with more and more ports. This also indirectly confirms that a relatively small number of port selections can still yield a near-optimal solution.

\subsection{Computational Performance}\label{ssec:compute}
This subsection presents the computational performance evaluation of the designed instruction-driven FPGA-based accelerator with a customized ISA and a specialized micro-architecture. The experiments are conducted on a Xilinx Virtex-7 XC7V690T FFG1761-3 FPGA platform and the design is synthesized using Xilinx Vitis HLS version 2022.2. The FPGA-based GNN accelerator employs a 8-bit fixed-point data representation, which offers greater resource efficiency and lower power consumption compared to floating-point arithmetic, with only negligible loss in accuracy. Moreover, this format also improves memory interface efficiency by shortening the delay in accessing off-chip memory.

In the experiments, the FPGA-based accelerator utilizes $64$ bits of off-chip memory bandwidth, while the remaining bandwidth is reserved for wireless communication and signal processing modules. When the port selection is limited to a single option, the inference latency ranges from $392,636$ to $610,442$ clock cycles, corresponding to $3.926$ to $6.104$ milliseconds, under a target clock period of $10~{\rm ns}$.


With fewer than four port selections, the inference latency increases only marginally. For example, when there are four port selections, the inference latency ranges from $399,418$ to $622,246$ clock cycles, corresponding to $3.994$ to $6.222$ milliseconds under a target clock period of $10~{\rm ns}$. It can be observed that the inference latency for four concurrent tasks is close to that of a single task. This is primarily due to the efficient reuse of neural network parameters already loaded into on-chip memory, as well as the full utilization of the computing cores. More specifically, since the GNN workload is memory-bound, the compute latency is significantly lower than the memory access latency. Only when the number of concurrent tasks reaches four does the compute latency become comparable to the memory access latency. On the other hand, the slight increase in inference latency from single-task to four-task execution mainly results from the small increase in the amount of input data to the model. However, as the number of port selections further increases, the inference latency rises significantly. This is mainly because the workload shifts from being memory-bound to compute-bound, as more intensive computations are required to handle multiple inference tasks. Additionally, the increased number of selected ports leads to a larger volume of intermediate data, potentially exceeding the capacity of on-chip storage.

These results demonstrate the effectiveness of the proposed task scheduling algorithm for the FPGA-based accelerator, achieving up to an approximately $4\times$ reduction in latency with only a modest increase in memory overhead. It is worth noting that although the current inference latencies exceed typical channel coherence time, they represent baseline performance on FPGAs used primarily for prototyping. Future advancements in model compression, sparse computation, and dedicated application-specific integrated circuit (ASIC) implementations are expected to further reduce inference latency, thereby enhancing the practicality of the proposed approach.

\section{Conclusions}\label{sec:conclude}
This paper proposed a hardware-software co-design approach for the joint optimization of beamforming and port selection in FASs. Using FA-enhanced downlink multi-cell MIMO networks as an example, simulation results demonstrated that the proposed GNN-RPS algorithm significantly outperforms traditional schemes including MMSE, ZF, and MRT. In addition, experimental results confirmed that the instruction-driven FPGA-based accelerator achieves low inference latency, ranging from $3.926$ to $6.104$ milliseconds, under a $10~{\rm ns}$ clock period with 8-bit fixed-point arithmetic for single-port selection. Thanks to the proposed task scheduling algorithm, the accelerator can efficiently process up to four port selections in parallel with only a marginal increase in latency. These findings highlighted the effectiveness and efficiency of the proposed hardware-software co-design approach for real-time intelligent beamforming in FASs.

\ifCLASSOPTIONcaptionsoff
  \newpage
\fi


\begin{thebibliography}{1}
%
\bibitem{New2025Tutorial}
W. K. New {\em et al.}, ``A tutorial on fluid antenna system for 6G networks: Encompassing communication theory, optimization methods and hardware designs,'' \emph{IEEE Commun. Surv. Tuts.}, early access, \url{doi: 10.1109/COMST.2024.3498855}, 2024.
\bibitem{Wong2020Fluid}
K.-K. Wong, K.-F. Tong, Y. Zhang, and Z. Zheng, ``Fluid antenna system for 6G: When Bruce Lee inspires wireless communications,'' {\em Elect. Lett.}, vol. 56, no. 24, pp. 1288--1290, Nov. 2020.
\bibitem{Wong2022Bruce}
K.-K. Wong, K.-F. Tong, Y. Shen, Y. Chen, and Y. Zhang, ``{Bruce Lee}-inspired fluid antenna system: Six research topics and the potentials for {6G},'' {\em Frontiers Commun. Netw.}, vol. 3, Mar. 2022, Art. no. 853416.

\bibitem{Sanayei2004Antenna}
S. Sanayei and A. Nosratinia, ``Antenna selection in MIMO systems,'' \textit{IEEE Commun. Mag.}, vol. 42, no. 10, pp. 68--73, Oct. 2004.
\bibitem{MIMO2004Molisch}
A. F. Molisch and M. Z. Win, ``MIMO systems with antenna selection,'' \textit{IEEE Microwave Mag.}, vol. 5, no. 1, pp. 46--56, Mar. 2004.

\bibitem{I22_wong2020perflim}
K. K. Wong, A. Shojaeifard, K. F. Tong, and Y. Zhang, ``Performance limits of fluid antenna systems,'' {\em IEEE Commun. Lett.}, vol. 24, no. 11, pp. 2469--2472, Nov. 2020.
\bibitem{I20_wong2021FAS}
K. K. Wong, A. Shojaeifard, K. F. Tong, and Y. Zhang, ``Fluid antenna systems,'' {\em IEEE Trans. Wireless Commun.}, vol. 20, no. 3, pp. 1950--1962, Mar. 2021.

\bibitem{I24_shen2024design}
Y. Shen {\em et al.}, ``Design and implementation of mmWave surface wave enabled fluid antennas and experimental results for fluid antenna multiple access,'' {\em arXiv preprint}, \url{arXiv:2405.09663}, May 2024.
\bibitem{I26_zhang2024pixel}
J. Zhang {\em et al.}, ``A novel pixel-based reconfigurable antenna applied in fluid antenna systems with high switching speed,'' {\em IEEE Open J. Antennas \& Propag.}, vol.~6, no.~1, pp.~212-228, Feb. 2025.
\bibitem{Liu-2025arxiv}
B. Liu, K. F. Tong, K. K. Wong, C.-B. Chae, and H. Wong, ``Be water, my antennas: Riding on radio wave fluctuation in nature for spatial multiplexing using programmable meta-fluid antenna,'' {\em arXiv preprint}, \url{arXiv:2502.04693}, 2025.
\bibitem{Lu-2025}
W.-J. Lu {\em et al.}, ``Fluid antennas: Reshaping intrinsic properties for flexible radiation characteristics in intelligent wireless networks,'' {\em IEEE Commun. Mag.}, vol. 63, no. 5, pp. 40--45, May 2025.

\bibitem{Chen2025Joint}
Y. Chen \textit{et al.}, ``Joint beamforming and antenna design for near-field fluid antenna system," \textit{IEEE Wireless Commun. Lett.}, vol. 14, no. 2, pp. 415--419, Feb. 2025.
%
\bibitem{Zhou2024Fluid}
L. Zhou \textit{et al.},  ``Fluid antenna-assisted ISAC systems," \textit{IEEE Wireless Commun. Lett.}, vol. 13, no. 12, pp. 3533--3537, Dec. 2024.
%
\bibitem{Qin2024Antenna}
H. Qin \textit{et al.}, ``Antenna positioning and beamforming design for fluid antenna-assisted multi-user downlink communications," \textit{IEEE Wireless Commun. Lett.}, vol. 13, no. 4, pp. 1073--1077, Apr. 2024
%
\bibitem{Zou2024Shifting}
J. Zou \textit{et al.}, ``Shifting the ISAC trade-off with fluid antenna systems,'' \textit{IEEE Wireless Commun. Lett.}, Vol. 13, No. 12, pp. 3479--3483, Dec. 2024.
%
\bibitem{Efrem2024Transmit}
C. N. Efrem and I. Krikidis, ``Transmit and receive antenna port selection for channel capacity maximization in fluid-MIMO systems," \textit{IEEE Wireless Commun. Lett.}, vol. 13, no. 11, pp. 3202--3206, Nov. 2024
%
\bibitem{Wang2024Learning}
C. Wang {\em et al.}, ``Fluid antenna system liberating multiuser MIMO for ISAC via deep reinforcement learning,'' {\em IEEE Trans. Wireless Commun.}, vol.~23, no.~9, pp. 10879--10894, Sept. 2024.
%
%
\bibitem{Jiang2021Learning}
T. Jiang, H. V. Cheng, and W. Yu, ``Learning to reflect and to beamform for intelligent reflecting surface with implicit channel estimation,'' \textit{IEEE J. Sel. Areas Commun.}, vol. 39, no. 7, pp. 1931--1945, Jul. 2021.
%
\bibitem{Chen2024Distributed}
C. Chen \textit{et al.}, ``A distributed machine learning-based approach for IRS-enhanced cell-free MIMO networks,'' \textit{IEEE Trans. Wireless Commun.}, vol. 23, no. 5, pp. 5287--5298, May 2024.
%
\bibitem{Xu2025TVT}
S. Xu, G. Chen, Y. Ma and R. Tafazolli, ``Distributed hybrid beamforming for downlink multi-user space-MIMO communications,'' \textit{IEEE Trans. Veh. Tech.}, \url{doi: 10.1109/TVT.2025.3572224}, 2025.
%
\bibitem{Li2024GNN}
Y. Li \textit{et al.}, ``GNN-based beamforming for sum-rate maximization in MU-MISO networks," \textit{IEEE Trans. Wireless Commun.}, vol. 23, no. 8, pp. 9251--9264, Aug. 2024.
%
\bibitem{Mishra2024Graph}
S. Mishra, L. Salaun, H. Yang and C. S. Chen, ``Graph neural network aided power control in partially connected cell-free massive MIMO,'' {\em IEEE Trans. Wireless Commun.}, vol. 23, no. 9, pp. 12412--12423, Sept. 2024.
%
\bibitem{Li2024MaxMin}
Y. Li \textit{et al.}, ``GNN-enabled max-min fair beamforming,'' \textit{IEEE Trans. Veh. Tech.}, vol. 73, no. 8, pp. 12184--12188, Aug. 2024.
%
\bibitem{Xu2023Distributed}
X. Xu, Y. Liu, Q. Chen, X. Mu and Z. Ding, ``Distributed auto-learning GNN for multi-cell cluster-free NOMA communications,'' \textit{IEEE J. Sel. Areas Commun.},  vol. 41, no. 4, pp. 1243--1258, Apr. 2023.
%

%
\bibitem{Xu2015GSM}
S. Xu, Y. Du, G. Chen, and R. Tafazolli, ``GSM: A GNN-based space-MIMO framework for direct-to-cell communications,'' {\em arXiv preprint}, \url{arXiv:2412.07555}, 2024.
%
\bibitem{Wei2018TGPA}
X. Wei \textit{et al.}, ``TGPA: Tile-grained pipeline architecture for low latency CNN inference,'' in \textit{Proc. Int. Conf. Comput. Aided Design (ICCAD)}, 5-8 Nov. 2018, San Diego, CA, USA.
\bibitem{Zhang2020DNNExplorer}
X. Zhang \textit{et al.}, ``DNNExplorer: A framework for modeling and exploring a novel paradigm of FPGA-based DNN accelerator,'' in \textit{Proc. Int. Conf. Comput. Aided Design (ICCAD)}, 2-5 Nov. 2020, San Diego, CA, USA.
%
\bibitem{Genc2021Gemmini}
H. Genc, S. Kim, A. Amid, \textit{et al.}, ``Gemmini: Enabling systematic deep learning architecture evaluation via full-stack integration,'' in \textit{Proc. Design Autom. Conf. (DAC)}, 5-9 Dec. 2021, San Francisco, CA, USA.
\bibitem{Yu2019OPU}
Y. Yu, C. Wu, T. Zhao, K. Wang and L. He, ``OPU: An FPGA-based overlay processor for convolutional neural networks,'' \textit{IEEE Trans. Very Large Scale Integr. Syst.}, vol. 28, no. 1, pp. 35--47, Jan. 2020.
\bibitem{Yu2020Light-OPU}
Y. Yu, T. Zhao, K. Wang, and L. He, ``Light-OPU: An FPGA-based overlay processor for lightweight convolutional neural networks,'' in \textit{Proc. Int. Symp. Field-Programmable Gate Arrays (FPGA)}, pp. 122--132, 23-25 Feb. 2020, Seaside, CA, USA.
\bibitem{Tang2024Grapht-OPU}
E. Tang \textit{et al.}, ``Graph-OPU: A highly flexible FPGA-based overlay processor for graph neural networks,'' \textit{ACM Trans. Reconfigurable Technol. Syst.}, vol. 17, no. 4, pp. 1--33, Nov. 2024.
%
\bibitem{New2024information}
W. K. New, K.-K. Wong, H. Xu, K.-F. Tong and C.-B. Chae, ``An information-theoretic characterization of MIMO-FAS: Optimization, diversity-multiplexing tradeoff and $q$-outage capacity,'' {\em IEEE Trans. Wireless Commun.}, vol. 23, no. 6, pp. 5541--5556, Jun. 2024.
%

%
\bibitem{Xu2024Intelligent}
S. Xu \textit{et al.}, ``Intelligent reflecting surface enabled integrated sensing, communication and computation,'' \textit{IEEE Trans. Wireless Commun.}, vol. 23, no. 3, pp. 2212--2225, Mar. 2024.
%
\bibitem{Du2025Intelligent}
Y. Du, S. Xu, G. Zhang, B. Wu and J. Zhang, ``Intelligent reflecting surface backscatter downlink multi-user communications with radar sensing,'' \textit{IEEE Trans. Veh. Technol.}, vol. 74, no. 5, pp. 8351--8356, May 2025.

\bibitem{Zhang2015Optimizing}
C. Zhang \textit{et al.}, ``Optimizing FPGA-based accelerator design for deep convolutional neural networks,'' in \textit{Proc. Int. Symp. Field-Programmable Gate Arrays (FPGA)}, pp. 161--170, 22-24 Feb. 2015, Monterey, CA, USA.
\end{thebibliography}
\end{document}